\documentclass[12pt]{article}

\usepackage{graphicx}
\usepackage[english]{babel}

 \font\gotb eufm10 scaled \magstep1
 
\newcommand{\bb}{\bibitem}
\newcommand{\cc}{\cite}
\newcommand{\vp}{\varphi}
\newcommand{\sss}{\sigma}
\newcommand{\rv}{{\bf r}}
\newcommand{\Rv}{{\bf R}}
\newcommand{\nv}{{\bf n}}

\newcommand{\va}{\vartheta}
\newcommand{\T}{\Theta}

\newcommand{\al}{\alpha}
\newcommand{\Om}{\Omega}
\newcommand{\ee}{\varepsilon}
\newcommand{\lt}{\left}
\newcommand{\rt}{\right}
\newcommand{\lll}{\lambda}
\newcommand{\F}{{\cal F}}

\newcommand{\s}{\hat S}
\newcommand{\I}{\hat I}
\newcommand{\A}{\hat A}
\newcommand{\B}{\hat B}
\newcommand{\U}{\hat U}

\newcommand{\f}{\phi}

\newcommand{\ttau}{\hbox{\boldmath$\tau$}}
\newcommand{\AAA}{\hbox{\gotb A}}
\newcommand{\HHH}{\hbox{\gotb H}}
\newcommand{\QQ}{\hbox{\gotb Q}}
\newcommand{\RR}{\hbox{\gotb R}}

\newcommand {\QQQ}{\hbox {\gotb Q}_{\xi}}
\newcommand {\vx}{\vp_{\xi}}
\newcommand {\ve}{\vp_{\eta}}
\newcommand{\bea}{\begin{eqnarray} \label}
\newcommand{\eeq}{\end{equation}}
\newcommand{\beq}{\begin{equation} \label}
\newcommand{\eea}{\end{eqnarray}}
\newcommand{\eeaa}{\end{eqnarray*}}
\newcommand{\beaa}{\begin{eqnarray*}}
\newcommand{\nn}{\\ \nonumber}
\newcommand{\rr}[1]{(\ref{#1})}

\newcommand{\post}{{\sc Postulate}}

 \author{\it D.A.Slavnov }

\title{Macroscopic Simulation of Violation of Bell's Inequality
\thanks{Physics of Particles and Nuclei, 2010, Vol. 41, No. 5, pp. 766-777.} }

   \date{}

\begin{document}
\maketitle

\begin{center} {\it  Department of Physics, Moscow State
University,\\  Moscow 119899, Russia. E-mail:
slavnov@goa.bog.msu.ru }
 \end{center}

 \begin{abstract}

A macroscopic quantum model of a two-level system (the analogue of
a half-spin particle) is described. The model is employed for
simulating not only the system under study, but the measurement
process as well. Single- and two-particle state models of a
quantum system are constructed. The Einstein-Podolsky-Rosen
paradox and Bell's inequality are discussed within the framework
of the model.
\end{abstract}


\section{INTRODUCTION}

Bohr's collaborator A. Petersen cites a statement \cc{pet} he had
once heard from Bohr while discussing problems in quantum
mechanics: "There is no quantum world. There only is an abstract
physical quantum description."

Although this assertion is in no way obvious, we take it to serve
us in describing a physical model that possesses, on one hand,
typical quantum properties and consists of macroscopic components,
on the other.

Such a model can be quite useful in studying problems that still
provoke considerable controversy in the physics community. These
include, e.g., the problem of locality in quantum measurements,
the causality problem, the existence of a physical reality that
may affect results of measurement, etc.

According to widespread belief, the main feature of quantum
systems is their specific quantum dynamics. In the latter, a
significant role is played by processes in which the components of
a physical system exchange portions of action comparable to the
atomic unit of action $\hbar$.

In contrast to this point of view, we try to show in this work
that quantum dynamics, which determines the interactions among
different parts of a system studied, is in many cases inessential
for the onset of typical quantum features, whereas the
peculiarities of the interaction between a quantum system explored
and an external system, which serves the purpose of either the
formation of the state of a system under study or proper
measurement, are important.

Since the result of such an interaction depends not only on the
characteristics of a system under investigation, but also on the
properties of an external system, a likelihood exists that results
will be obtained that are typical of quantum phenomena even in the
case of macroscopic dimensions of a system examined only by
employing appropriate parameters of the external system. We shall
try to implement this program for a two-level physical system that
is analogous to a half-spin particle.

Further consideration will be carried out within the framework of
a special version of the algebraic approach to the quantum theory,
which is described in \cc{slav1,slav2}. In the latter references,
a phenomenological basis for the postulates of this approach is
given, along with a description of how to develop the standard
mathematical apparatus of quantum mechanics starting from these
postulates.

This approach has a number of distinctive features that will be
very useful in what follows. First, quantum and classical systems
are considered from a unified point of view and, moreover, the
quantum system can be regarded as a set of open {\it classical}
subsystems. Second, the dynamics is not fixed in this approach.
Therefore, it is equally applicable to both quantum and classical
systems. Third, the mathematical apparatus of the approach is
constructed as a mathematical description of the physical
measurement process. Recall that, within the traditional approach,
one tries (not very successfully) to develop a theory of quantum
measurements that is consistent with the a priori mathematical
apparatus of quantum mechanics.

\section{BASICS OF THE APPROACH}

In this section the fundamental axioms of the approach will be
listed and briefly commented on. More detailed accounts and
justifications can be found in \cc{slav1,slav2}.

Within the framework of the algebraic approach, the primary
building blocks of the quantum theory are observables (see,
e.g.,~\cc{emch,hor,blot}). The latter are characteristics of a
physical system that can be given numerical values by using
certain measurement processes. Upon choosing a specific system of
units, all the observables can be considered dimensionless. The
main property of the observables is that they can be multiplied by
real numbers and form products between and add up to each other.
Therefore, the following postulate is adopted in the algebraic
approach.

\

\post{} 1. Observables $\A$ of a physical system are Hermitian
elements of some $C^*$-algebra \AAA, ($\A\in \AAA_+ \subset \AAA$,
$\A^*=\A$).

\

Here~$\AAA_+$ is the set of observables.

\

Recall that algebra is a set that is a linear space with a defined
operation of multiplication of the elements. The algebra \AAA{} is
called $C^*$-algebra (see, e.g.,~\cc{dix}), if a conjugation (an
involution) operation $\U\to\U^*$ $(\U\in\AAA,\quad \U^*\in\AAA)$
is defined, and the norm of an arbitrary element $\U$ satisfies
the following condition $\|\U\U^*\|=\|\U\|^2$.

The fundamental difference between quantum and classical systems
lies in the fact that any pair of observables in a classical
system can be measured in a compatible manner, which means that
multiple measurements of two observables $\A$ and  $\B$ will give
rise to identical results for each of them regardless of the
sequence of measurements undertaken. For quantum systems, only
certain groups of observables can be measured in such a way. The
observables are called compatible if they belong to the same
group. Mark each such group with the index~$\xi$ that takes its
values in the set $\Xi$ and distinguishes one such group from
another. The same observable $\A$ can belong to many
groups~$\QQQ$. Each group~$\QQQ$ can be treated as a set of
observables of some classical subsystem of a quantum system
studied. As a consequence, the following two postulates are
adopted.

\

\post{} 2. A set of compatible (simultaneously measurable)
observables is a maximal real associative commutative
subalgebra~$\QQQ$ of the algebra~\AAA \quad ($\QQQ\subset\AAA_+$).

\

\post{} 3. The state of a classical subsystem, with its
observables being the elements of subalgebra $\QQQ$, is described
by the character $\vx(\cdot)$ of this subalgebra.

\

The latter means that $\A\stackrel{\vx}{\longrightarrow}\vx(\A)$
($\A\in \QQQ$) is a homomorphism of the algebra $\QQQ$ into the
set of real numbers.

Let us now consider a quantum system to be a set of open classical
subsystems with their corresponding observables being the elements
of the corresponding subalgebra $\QQQ$. Bearing in mind that each
observable that belongs to the algebra \AAA{} is also associated
with some subalgebra $\QQQ$, we adopt the following postulate.

\

\post{} 4. The result of each individual measurement of the
observable of a physical system is defined by the elementary state
$\vp=[\vx]$ of this system.

\

Here $\vp=[\vx]$ is the set of characters of all subalgebras
$\QQQ$, with each of them represented by just one character
$\vx(\cdot)$  in $\vp=[\vx]$.

Instrument readings are due to interaction between an instrument
and a physical system. This result may depend on the
characteristics of both the system studied and the instrument. The
latter is highly undesirable when examining the system under
scrutiny. The problem of unification of the readings obtained from
different instruments is usually solved with the help of a
calibration procedure.

This procedure is schematically as follows. The instrument is
taken to be a template that provides reproducible measurement of
some observable $\A$. This instrument measures the observable $A$.
Hereinafter, the result of measurement is denoted by the same
symbol as an observable itself, but without a "hat." If the first
measurement was carried out in a reproducible manner, then the
duplicate measurement of the same observable by the instrument
subject to calibration should yield the same value. Unification of
readings of both the template and the instrument to be calibrated
works in this way.

In a quantum case, however, such a procedure is feasible only
within each group of compatible observables. Let us say that the
instruments allowing for such measurements within a group $\QQQ$,
are related to the $\xi$-type. In the general case, reading
unification for different-type instruments is impossible to
realize. For a quantum system, even in the ideal case, dependence
of the result of measurement on the type of instrument cannot be
excluded.

If the observable $\A$ belongs simultaneously to two different
subalgebras ($\A\in \QQ_{\eta}\cap\QQ_{\eta'} \quad \eta, \eta'\in
\Xi$), then the different type ($\eta$ and $\eta'$) instruments
can be used to measure it and the equality
 \beq{1}
\vp_{\eta}(\A)=\vp_{\eta'}(\A), \hbox{ of } \ve,\vp_{\eta'} \in
\vp=[\vx],\quad \eta\neq \eta'
 \eeq
may be violated (see \cc{slav1}).

If for some $\vp=[\vx]$ equality \rr{1} holds for all $\QQ_{\xi}$
containing the observable $\A$, then we say that the elementary
state $\vp=[\vx]$ is stable with respect to the observable $\A$.

If the elementary state $\vp=[\vx]$ of the system studied was
known to us, then we would be able to predict unambiguously the
result of measurement of any observable. However, this is
impossible. Only compatible observables can simultaneously be
measured, for example, those associated with the subalgebra
$\QQ_{\eta}$. Therefore, out of all the characters $\vx$ inherent
to the elementary state $\vp$, only the functional
$\vp_{\eta}(\cdot)$. can experimentally be determined. In this
connection, it is convenient to introduce the notion of
$\ve$-equivalence.

Two elementary states $\vp$ and $\vp'$ are called $\ve$-equivalent
if for all $\A\in\QQ_{\eta}$ the following equality holds:
$$ \ve(\A)=\vp'_{\eta}(\A), \mbox{ where } \ve\in \vp, \;
\ve'\in\vp'. $$
 Accordingly, what we can at most learn of the
elementary state in experiment is that it belongs to the certain
equivalence class $\{\vp\}_{\ve}$.  If the observable   $\A$
belongs to the subalgebra $\QQ_{\eta}$ and elementary
$\vp\in\{\vp\}_{\ve}$ is stable with respect to the observable
$\A$, then as a result of measurement we definitely obtain
$A=\ve(\A)$.  If $\A\neq\QQ_{\eta}$, then it is impossible to
predict definitely the result of measurement, since these may
differ for different elementary states $\vp\in\{\vp\}_{\ve}$.

To put it differently, the equivalence class has the physical
properties that are ascribed to the quantum state in the
traditional convenient approach. Therefore, the set of equivalent
elementary states stable with respect to all the observables
 $\A\in \QQ_{\eta}$ is identified with the
quantum state
$$ \Psi_{\ve}\equiv \{\vp\}_{\ve}$$

The elementary state meets the requirements for the elementary
events in Kolmogorov probability theory \cc{kol} ((see also
\cc{nev}). Namely, only one elementary event is realized in each
test. Different elementary events are mutually exclusive.
Therefore, there is no need to devise some artificial quantum
probability theory; instead, the well-developed classical one can
be used. As the probabilistic properties of a quantum system are
fixed by its quantum states, it is natural to adopt the following
postulate.

\

\post{} 5. The equivalence class $\{\vp\}_{\ve}$ can be equipped
with the structure $(\Om,\F, P)$ of the probability space.

\

Here $\Om $ is the space of elementary events for which it is in
general impossible to define the probabilistic measure $P(F)$ in
Kolmogorov probability theory. In order to define the latter, it
is necessary to render the space $\Om $ measurable. To this end,
besides elementary events, the so-called (probabilistic) events
$F$. should be introduced. These are subsets of the $\Om $ set for
which the probabilistic measure $P(F)$ can be defined. It is
assumed that the event $F$ has occurred once an elementary event
belonging to the subset $F$ is realized. The subsets $F$ must be
the elements of so-called Boolean $\sss$-algebra $\F$. This means
that the set $\F$ is equipped with three algebraic operations: the
union of subsets $F$, their intersection, and the complement of
each subset to the full set $\Om$. The set  $\F$ must necessarily
include the set $\Om$ itself along with the empty set $\emptyset$.
In addition, this set $\F$ must be closed with respect to the
complement operation and a countable number of unions and
intersections. The latter property means that, under any of those
operations, we obtain an element of the same set. The space $\Om $
equipped with the described $\sss$-algebra $\F$ is called
measurable space. The choice of certain Boolean algebra $\F$
corresponds in terms of physics to the choice of instrument type.
The probabilistic measures $P(F)$ are defined only for the events
$F\in\F$.

A distinctive feature of the quantum measurements is that it is
possible to use simultaneously only those instruments which allow
for measurements of compatible observables, i.e., those which
belong to a certain $\xi$-type. There is certain Boolean algebra
$\F_{\xi}$. that corresponds to each type of such instruments. It
is essential that, in a quantum case, there is no Boolean algebra
$\F_0$ with the following properties: for all $F\in\F_0$ there are
probabilistic measures $P(F)$ and the algebras $\F_{\xi}$ are
subalgebras of the algebra $\F_0$.

Postulate 5 allows defining the average value of the observable
$\A$ in a quantum state $\Psi_{\eta}$ using the probabilistic
average over a space $\{\vp\}_{\ve}$ of elementary (states)
events:

 \beq{2}
 \Psi_{\ve}(\A)=\int_{\vp\in\{\vp\}_{\ve}}\,P_{\A}(d\vp)\,\vx(\A).
 \eeq
 Here
 $ P_{\A}(d\vp)=P(\vp:\vx(\A)\leq A+dA)-P(\vp:\vx(\A)\leq A), \quad
 \A\in\QQQ, \quad \vx\in\vp\in\{\vp\}_{\ve}.$

In order for formula \rr{2} to define the quantum average, the
probabilistic measure $P_{\A}(\vp)$ must satisfy the following
postulates.

\

\post{} 6. The functional $\Psi_{\ve}(\A)$ is linear over the
algebra \AAA.

\

\post{} 7. The functional $\Psi_{\ve}(\A)$ does not depend on the
particular choice $\xi$.

\

The latter statement is to be understood as follows. The
observable $\A$ can simultaneously belong to several subalgebras
$\QQ_{\xi_1},\,\QQ_{\xi_2},...$. For all $\xi_1,\xi_2,...$ formula
\rr{3} must define the same functional. Since $\vx(\cdot)$ is the
character of the subalgebra $\QQQ$, the functional $\Psi_{\ve}$
will automatically be positive and normalized to unity, i.e.,
$\Psi_{\ve}(\I)=1$, where $\I$ is the identity element of the
algebra~\AAA.

With the $C^*$-algebra $\AAA$ and the linear positive normalized
functional $\Psi_{\ve}(\cdot)$ defined over this algebra, we can
construct a representation of the algebra $\AAA$ by using the
canonical Gelfand-Naimark-Segal construction (see, e.g.,
\cc{naj,emch}). In other words, we can construct Gilbert space
$\HHH$, in which there is an operator $\Pi(\A)$ acting over a
space $\HHH$ that corresponds to each element $\A\in\AAA$, while
the expectation value $\langle\Phi|\Pi(\A)|\Phi\rangle$, where
$|\Phi\rangle\in\HHH$ is the corresponding vector in Gilbert space
--- to the quantum average $\Psi_{\ve}(\A)$. This is the way the standard
mathematical apparatus of quantum mechanics is reproduced.

From the point of view of quantum calculations, the mathematical
apparatus based on Gilbert space usually turns out to be more
convenient. However, it bears no clear physical interpretation
and, therefore, is poorly adapted for macroscopic simulation of
quantum processes. In contrast, the formalism based on the
elementary state is applicable to both quantum and classical
systems. That is why it is convenient for such a simulation to be
performed.

\section{A TWO-LEVEL SYSTEM}

In this section we discuss an application of basics given in the
previous section to a certain quantum model. For the latter, let
us take a two-level system analogous to a half-spin particle. In
what follows the spin terminology will be used for the model
considered. At the same time, we emphasize that this does not
imply that the dynamics of the system under study is identical to
that of a half-spin particle.

The observables of the two-level quantum system can be represented
by the Hermitian matrices $2\times 2$. In this case, the set of
all the matrices of the type

$$ \A=\lt[\begin{array}{cc}
  a & b \\
  c & d
\end{array}\rt], $$
can be considered to be the algebra \AAA, in which algebraic
operations coincide with the corresponding matrix ones.

For such a system, it is not difficult to construct all the
elementary states. Let $\A$ be the Hermitian matrix; then,
$a^*=a$, $d^*=d$, and $c=b^*$. Any matrix of such a type can be
represented in the form
 \beq{029}
 \A=g_0\I+g\,\hat\tau(\rv).
 \eeq
Here $\rv$ is the three-dimensional unit vector, $\tau_i$ are the
Pauli matrices, and $\hat\tau(\rv)=(\ttau\cdot\rv)$. In order for
formula \rr{029} to hold, it is required to set
 \bea{03}
& g=\lt((a-d)^2/4+b\,b^*\rt)^{1/2}, \quad
 g_0=(a+d)/2,\nn & r_1=(b+b^*)/(2g), \quad
 r_2=(b-b^*)/(2ig), \quad r_3=(a-d)/(2g).
 \eea

It is obvious that $\hat\tau(-\rv)=-\hat\tau(\rv)$. For
$\rv'\ne\pm\rv$, the commutator of matrices $\hat\tau(\rv)$ and
$\hat\tau(\rv')$ is nonzero. Therefore, each (up to the sign)
matrix $\hat\tau(\rv)$ is the generatrix for the real maximal
commutative subalgebra $\QQ_{\rv}$. Since
$\hat\tau(\rv)\hat\tau(\rv)=\I$, then the element spectrum
$\hat\tau(\rv)$ consists of two points: $\pm1$.

Let $\vp^{\al}=[\vp^{\al}_{\rv}]$ be an elementary state. Here
$\vp^{\al}_{\rv}$ is the character of the subalgebra $\QQ_{\rv}$
and the index $\al$ distinguishes one elementary state from the
other. Consider the function $f^{\al}(\rv)$ such that
$f^{\al}(-\rv)=-f^{\al}(\rv)$ and for each $\rv$ the function
takes either $+1$ or $-1$, with the index $\al$ once again
distinguishing the functions from each other. Then we can
apparently assume that
$\vp^{\al}_{\rv}(\hat\tau(\rv))=f^{\al}(\rv)$. Taking into account
that $\hat\tau(\rv)$ is the generatrix of the subalgebra
$\QQ_{\rv}$, we obtain
 $$\vp^{\al}_{\rv}(\A)=g_0(\A)+g(\A)f^{\al}(\rv)$$
we obtain $\A\in\QQ_{\rv}$.

Note that for the quantum system in question each observable $\A$
(aliquant to $\I$) belongs to the same maximal subalgebra
$\QQ_{\rv}$.  Of course, this in general is not correct. By taking
advantage of the mentioned feature inherent to the system at hand,
we can represent the elementary state in the form of a universal
functional defined over the entire set $\AAA_+$:
 \beq{131}
\vp^{\al}(\A)=g_0(\A)+g(\A)f^{\al}(\rv(\A))
 \eeq
for any observable $\A$. Here, not only $g$ and $g_0$, but also
$\rv$ should be considered as functions of $\A$ (see
formula~\rr{03}). This functional is explicitly extended over the
whole algebra \AAA. This example is evidence that the proof by von
Neumann~\cc{von} of nonexistence of hidden parameters is
inapplicable to elementary events. Let us draw attention to the
fact that the functional $\vp^{\al}(\A)$ defined by
formula~\rr{131} is nonlinear. In a general case, the elementary
state can also be formally represented in the form of a nonlinear
functional defined over the whole algebra \AAA. However, this
functional will in general be ambiguous, because the same
observable $\A$ may belong to several maximal commutative
subalgebras $\QQQ$.

From formula \rr{131} it follows that, in order to define the
functional $\vp^{\al}(\A)$, it is sufficient to define the
function  $f^{\al}(\rv)$. This means that each function
$f^{\al}(\rv)$ is in one-to-one correspondence with a certain
elementary state. Let us designate the value of the function
$f^{\al}(\rv)$ equal to $+1$ with a black dot and $-1$ with a
white one. Then, each of such elementary states can be visualized
as a sphere of unit radius the surface of which is spotted by
black and white dots, with those of different colors occupying the
ends of each diameter. By using these spheres, we will be able to
model elementary states of a half-spin particle.

Let the system be in a definite elementary state to which the
sphere (the elementary state sphere (ESS)) described above
corresponds. To define the outcome of measurement of spin
projection onto a definite direction, it is required to draw a
unit vector $\rv$ from the center of the ESS along that particular
direction. If this vector encounters a black (white) dot on a
sphere, then the result of measurement will be $+1/2$ ($-1/2$).

Thus, consider a half-spin quasiparticle the elementary states of
which can be described by pointwise painted ESSs. Very close
points may have different colors. Therefore, an infinitesimal
rotation of the instrument may give rise to a drastic change in
the result registered. At the same time, experience shows that, if
we study some quantum state, i.e., some set of ESSs, then on
average a small rotation of the instrument causes a mild change in
the result. This means that, from the point of view of finding
average values, we can substitute a set of pointwise painted
spheres in the quantum state for a set of continuously painted
ones. Of course, both the coloring and ensemble of such spheres
should be selected properly.

Let us assume that all the spheres have identical coloring and
differ solely in their orientations. The sphere of the north
magnetic pole (+1) is painted black, and that of the south
magnetic pole $(-1)$ white. Intermediate areas are gray, their
darkness varying according to the law

  \beq{3}
\rho=({\rv\Rv}).
 \eeq
Here $\rv$ is the radius-vector drawn toward a current gray dot,
while $\Rv$ denotes the radius-vector drawn toward the north
magnetic pole.

Transition to gray-painted spheres can resolve the problem of a
smooth change of registered average values caused by the
instrument rotation, but, at the same time, gives rise to the new
problem. Each measurement must yield either $+1/2$ or $-1/2$ for
the spin projection, not some intermediate "gray" value.

To overcome this obstacle, assume that the elementary state of a
particle is characterized by a multilayer gray-painted sphere
(MGS), with each layer painted according to the law given in
\rr{3} and with different layers having different orientations.
The orientation of the $k$ layer is denoted by the vector
$\Rv^{(k)}$. In addition, assume that the elementary state is
characterized by the values of auxiliary variable $\hat \ee$. A
(failed) attempt to use such a variable was undertaken in
\cc{matz}. Let $\rv$ be the radius-vector drawn toward a current
dot on a sphere. Then, each value of $\rv$ corresponds to the
value $\ee(\rv)$ of the variable $\hat \ee$. of the variable
$\ee(\rv)$ take, independently of each other, random values from
within the interval

$$ -1/2< \ee^{(k)}(\rv) < +1/2 $$
and
$$ \ee^{(k)}(-\rv)=-\ee^{(k)}(\rv)$$.

Now, out of all such MGSs, select a subset $\Upsilon$ that
consists of those $\ee(\rv)$ of which related to each layer
satisfy one of the following conditions:

\beq{4}
 |\Rv^{(k)}\rv+\ee^{(k)}(\rv)|>1/2,
 \eeq

or

\beq{4a}
 |\Rv^{(k)}\rv+\ee^{(k)}(\rv)|<1/2.
 \eeq
It is easy to verify that such functions (discontinuous in
general) exist almost for each $\rv$ at an arbitrary orientation
of the layer. It is also clear that the subset $\Upsilon$ is
invariant with respect to the spherical transformations. Below, we
shall deal with only those MGSs that belong to a subset
$\Upsilon$.

The uppermost layer that satisfies condition \rr{4} is called
active. Now the associate ESS with each MGS equipped with the
functions $\ee(\rv)$ according to the following rule. If the
active layer of the MGS considered has the number $k$, then the
radius-vector r of the corresponding ESS is set against either a
black dot $(j=+1)$ if $\Rv^{(k)}\rv>0$  or a white dot $(j=-1)$ or
a white dot $\Rv^{(k)}\rv<0$. Recall that the dot color on the ESS
describes the result of measurement of the spin projected onto the
corresponding direction when a particle is in the elementary state
under consideration. Accordingly, assume that the instrument that
measures the spin projection onto the direction of $\rv$ operates
as follows. First, it checks if the uppermost layer of the MGS
corresponding to the elementary event considered is active. If it
is, then the instrument registers the event according to the rule
outlined above. If it is passive, then the instrument fails to
register any event and proceeds to study the next layer in the
same fashion. This process continues until there is a definite
result registered. We shall see later that the probability of the
active layer having a number greater than k rapidly decreases with
increasing $k$. In this way, a definite result will be registered
for nearly every elementary event.

Having in mind the correspondence between the MGS and ESS, we
shall construct the quantum states we are interested in by using
sets of MGSs rather than ESSs.

The quantum state with the doubled spin projection (equal to +1)
onto the direction of $\nv$ corresponds to the set of ESSs, with a
black dot found at the end of the radius of each member sphere
directed along $\nv$. The set of MGSs ($\Upsilon_{\nv}$) that
satisfies the following conditions is associated with this quantum
state. The radius-vector $\Rv^{(k')}$ in each layer $k'$ pointing
toward the north magnetic pole belongs to the upper hemisphere
$\RR^+_{\nv}$ with a central direction $\nv$. The orientations of
the vectors $\Rv^{(k')}$ for different layers are random and
distributed over $\RR^+_{\nv}$ independently of each other. Each
MGS included into $\Upsilon_{\nv}$ is equipped with the functions
$\ee^{(k')}(\rv)$. The latter functions are random as well and
satisfy either condition \rr{4} or \rr{4a}.

Any individual MGS with given functions $\ee^{(k')}(\rv)$ define
some elementary state. In probability theory, the latter
corresponds to some elementary event. According to probability
theory \cc{kol}, it is not at all necessary that an elementary
event be associated with some probabilistic measure. The
probabilistic measures must correspond to (probabilistic) events
that are some subsets of the set of elementary events.

Therefore, the notion of probability will only be associated with
certain subsets of the set of elementary states. For every value
of $\rv$ we construct a proper system of such subsets, a proper
$\sss$-algebra $\F_{\rv}$. This is consistent with the quantum
case, in which there is proper $\sss$-algebra (see \cc{slav1}) for
each group of compatible observables. In the case at hand, such a
group consists of observables that are linear combinations of the
spin projection onto the direction of $\rv$ and unity.

The $\sss$-algebras $\F_{\rv}$ generated are defined by the
following conditions.
\begin{enumerate}
\item The vector  $\rv$ fixes $\sss$-algebra $\F_{\rv}$. The
remaining conditions fix the elements of this algebra.
 \item The
interval $(\ee,\ee+d\ee)$ is fixed, where the parameter $\ee$
satisfies inequalities $-1/2<\ee<+1/2$. Then, the interval itself
and its length will be denoted by a single symbol $d\ee$.
 \item  Either $\ee^{(k')}(\rv)\in d\ee$ or $-\ee^{(k')}(\rv)\in d\ee$.
  Both alternatives correspond to the same subset (one element of
$\sss$-algebra).
 \item The number $k$ of the active layer is fixed
 \item Each subset consists of all the MGSs the layers of
which satisfy one of the following conditions:
 \bea{5}
 |\Rv^{(k')}\rv+\ee^{(k')}(\rv)|&>&1/2 \qquad k'=k,\\ \label{6}
|\Rv^{(k')}\rv+\ee^{(k')}(\rv)|&\leq&1/2 \qquad k'\leq k \quad (k'
\mbox{ fixed}).
 \eea
 The conditions given in \rr{5} and \rr{6} correspond to different
  subsets.
 \item If the condition in \rr{5} is satisfied, the subsets
are distinguished by another feature: either $\Rv^{(k)}\rv>0$ or
$\Rv^{(k)}\rv<0$.
\end{enumerate}

It is easy to verify that, irrespective of $\rv$, any elementary
event from $\Upsilon_{\nv}$ belongs to any listed subset at some
$\ee$ and $k$.

Now construct the probabilistic measures for each such subset,
assuming that the random values of $\Rv^{(k')}$ are evenly
distributed on a hemisphere $\RR^+_{\nv}$.

Let $\rv$ and $ d\ee$ fixed and $k'=1$. The probability for
inequality \rr{5} to hold with the additional condition of either
$\Rv^{(k)}\rv>0$ ($j=+1$) or $\Rv^{(k)}\rv<0$ ($j=-1$) is
described by the following expression:

\bea{7} P_{\nv}^{(1)}(\rv,\ee,j) d\ee &=& d\ee\frac N2 \int d\Rv\;
\T(\Rv\nv)\Big[\T[j(\Rv\rv+\ee)-1/2]+\T[j(\Rv\rv-\ee)-1/2]\Big]
\nn &=& d\ee\frac N2 \int d\Rv\;
\T(j\Rv\nv)\Big[\T[\Rv\rv+\ee-1/2]+\T[\Rv\rv-\ee-1/2]\Big].
 \eea
Here $N$ is the normalization factor, $d\Rv=d\f\,d\va\sin\va$, and
$\T(x)$  is the Heaviside step function ($\T(x)=0$ for $x<0$,
$\T(x)=1$ for $x>0$). From \rr{7} it follows that

   \beq{8}
P_{\nv}^{(1)}(\rv,\ee)\equiv \sum_{j=\pm 1}
P_{\nv}^{(1)}(\rv,\ee,j)=N\pi.
 \eeq

The probability for inequality \rr{6}  to hold is given by

\bea{9} \tilde P_{\nv}^{(1)}(\rv,\ee) d\ee &= &d\ee\frac N2 \int
d\Rv\; \T(\Rv\nv)
\Big[\T[1/2+\Rv\rv+\ee]\T[1/2-\Rv\rv-\ee]\nn&+&\T[1/2+\Rv\rv-\ee]\T[1/2-\Rv\rv+\ee]\Big]
\nn &=& d\ee\frac N2 \int d\Rv\;
\Big[\T[1/2+\Rv\rv+\ee]\T[1/2-\Rv\rv-\ee]\Big]=N\pi\;d\ee.
 \eea

From \rr{8} and \rr{9}, we obtain $N=(2\pi)^{-1}$ and

 \beq{10}
\tilde P_{\nv}^{(1)}(\rv,\ee)=1/2.
 \eeq

Now consider the case $k'=2$. The second layer has the same
properties as the first one; at the same time, its orientation and
the corresponding function $\ee^{(k')}(\rv)$ are independent of
the first layer. As a result, we can repeat previous arguments for
the second layer. The only thing to take into account is that,
because of formula \rr{10}, the instrument will deal with the
second layer with a probability of 1/2. This means that for a
given $ \ee$ the probability of obtaining the number of the active
layer equal to 2 and the fixed value of $j$ as a result is
described by the following formula:

$$ P_{\nv}^{(2)}(\rv,\ee,j) =\frac 1{2\pi} \frac 1{2^2} \int d\Rv\;
\T(j\Rv\nv)\Big[\T[\Rv\rv+\ee-1/2]+\T[\Rv\rv-\ee-1/2]\Big],
 $$
while the probability for the active layer to have a number
greater than 2 is as follows:
$$\tilde P_{\nv}^{(2)}(\rv,\ee)=(1/2)^2.$$

Continuing this process we obtain that, for fixed  $\rv$, $d\ee$
and $j$, the probability for the active layer to have the number
$k$ is equal to

\beq{11} P_{\nv}^{(k)}(\rv,\ee,j) =\frac 1{2\pi} \frac 1{2^k} \int
d\Rv\; \T(j\Rv\nv)\Big[\T[\Rv\rv+\ee-1/2]+\T[\Rv\rv-\ee-1/2]\Big],
 \eeq
whereas the probability of finding an active layer with a number
greater than $k$ is represented by
 \beq{12} \tilde P_{\nv}^{(k)}(\rv,\ee)=(1/2)^k.
 \eeq

It is noteworthy that the latter probability depends on neither
$\rv$ nor $\ee$ and rapidly decreases with increasing $k$.
Formulas \rr{11} and \rr{12} describe a system of probabilistic
measures corresponding to the $\sss$-algebra chosen.

As $\ee^{(k')}$ depend on  $\rv$, the subsets of elementary states
corresponding to the same interval $ d\ee$ will be different for
different $\rv$, which means that different instruments correspond
to different $\sss$-algebras. In other words, different
instruments split the set of elementary events into subsets in
different ways. Only within a single such partition is it
meaningful to speak about the probability of a specific elementary
event falling into one such subset. On the other hand, it is
meaningless to compare the probabilities of the elementary events
falling into subsets that belong to different partitions, since
such subsets are elements of differen $\sss$-algebras. In their
turn, these $\sss$-algebras are not subalgebras of some universal
$\sss$-algebra to which some system of probabilistic measures
corresponds. This is a distinctive feature of the quantum
measurements (see \cc{slav1}).

From \rr{11} it follows that the probability of obtaining a fixed
value for $j$ at fixed $\rv$ and an arbitrary number of the active
layer and $\ee$ is given by the formula

\begin{eqnarray*}
 P_{\nv}(\rv,j)& = &\sum_{k=1}^{\infty} \int^{1/2}_{-1/2} d\ee P_{\nv}^{(k)}(\rv, \ee,j)
 \\&=& \frac 1 {2\pi} \int^{1/2}_{-1/2} d\ee \int d\Rv\;
\T(j\Rv\nv)\Big[\T[\Rv\rv+\ee-1/2]+\T[\Rv\rv-\ee-1/2]\Big] \\ &=&
\frac 1{\pi} \int d\Rv\; \T[j\Rv\nv](\Rv\rv)\T[\Rv\rv].
 \end{eqnarray*}

 It follows then that

 \beq{13}
\sum_j P_{\nv}(\rv,j)=\frac 1{\pi} \int d\Rv\;
(\Rv\rv)\T[\Rv\rv]=1.
 \eeq
and \beq{14}
 \sum_j j\;P_{\nv}(\rv,j)=\frac 1{\pi}\sum_j \int d\Rv\;\T[j\Rv\nv]
j(\Rv\rv)\T[\Rv\rv]=\int d\Rv\;\T[\Rv\nv] (\Rv\rv)=(\rv\nv).
 \eeq

 From \rr{13} and \rr{14}, we derive
 \beq{15}
 P_{\nv}(\rv,j)=\frac 12 \Big(1+j(\rv\nv)\Big).
 \eeq

 This distribution coincides with the quantum one.

\section{EPR PARADOX, BELL'S INEQUALITY}

In their famous work \cc{epr}, Einstein, Podolsky, and Rosen
formulated the principles to be satisfied by a complete physical
theory: (a) "each element of physical reality must have a
counterpart in a complete physical theory" and (b) "if we are able
with certainty (i.e., with a unit probability) to predict the
value of a physical quantity without perturbing a system, then it
means that there is an element in the real world that corresponds
to this quantity."

The standard mathematical apparatus of quantum mechanics does not
meet this requirement. An individual experiment happens to have no
adequate counterpart in this approach. This leads to paradoxical
conclusions if one works within the framework of the standard
formalism of quantum mechanics. One such conclusion is the
renowned Einstein-Podolsky-Rosen (EPR) paradox. In the original
work~\cc{epr} this paradox was formulated using an example with
coordinate and momentum measurements, whereas Bohm proposed a
simpler physical model~\cc{bom} in which the same old problem is
discussed using the example of measurement of spin projections
onto various directions. Here we shall be concerned with the Bohm
model.

It appears as follows. A zero-spin particle decays into two
half-spin particles $A$ and $B$, which are scattered far apart.
The spin state of this system is described by the following
formula of standard quantum mechanics:
 \beq{46}
 |\Psi\rangle=\frac{1}{ \sqrt{2}}\lt[|A_z^{(+)}\rangle|B_z^{(-)}\rangle
 -|A_z^{(-)}\rangle|B_z^{(+)}\rangle\rt],
 \eeq
where $|A_z^{(\pm)}\rangle$, $|B_z^{(\pm)}\rangle$ are the
eigenvectors of the operators of spin projections onto the $z$
axis with eigenvalues of $+1/2$ and $-1/2$, respectively. This is
the so-called entangled state, in which neither particle $A$, nor
$B$ has a certain value for the spin projection onto $z$ axis.

Once the $A$ and $B$ particles move far apart, the $B$ particle's
spin projection onto the $z$ axis is measured. Let the result of
this measurement be $+1/2$. Then, according to the postulate of a
quantum state collapse (projection principle), the state
$|\Psi\rangle$ is instantly replaced by the state
$|\tilde\Psi\rangle=-|A_z^{(-)}\rangle |B_z^{(+)}\rangle$. This
means that the subsequent measurement of the $A$ particle's spin
projection onto $z$ axis the value $-1/2$ will be obtained with
unit probability, thus describing the experiment very well.
However, from the physics point of view, this result seems to be
paradoxical.

Indeed, the following reasoning appears to be most natural: at the
moment of decay, the particles $A$ and $B$ acquire certain spin
projections onto the $z$  axis (opposite in sign), but until $B'$s
spin projection is measured we do not know their definite values.
Once $B'$s spin projection is measured, that of the $A$ particle
is automatically known. However, such an explanation contradicts
the general concept of the standard quantum mechanics.

 The point is that the same quantum state $|\Psi\rangle$ can be
represented in the form
 $$|\Psi\rangle=\frac{1}{ \sqrt{2}}\lt[|A_x^{(+)}\rangle|B_x^{(-)}\rangle
 -|A_x^{(-)}\rangle|B_x^{(+)}\rangle\rt],$$
 where the notation is the same as in formula~\rr{46},
except for the projection onto the  $x$ rather  than the $z$ axis.
Now we can repeat all the arguments given after formula~\rr{46},
with the $z$ axis replaced by the  $x$ one. As a result, we obtain
that, at the moment of decay, the particles must acquire definite
values for the spin projections onto the $x$ axis. However, the
observables corresponding to the spin projections onto the $z$ and
$x$ axes are mutually incompatible and, therefore, cannot
simultaneously have definite values, according to standard quantum
mechanics.

An alternative might look as follows. After decay the particles
$A$ and $B$ did not acquire definite values for the spin
projections onto any axis. Once those projections onto certain
axis were measured, they acquired definite values for the former.
It is not difficult to imagine that this mechanism is feasible for
the particle $B$ that interacted with the instrument. On the other
hand, the particle $A$ is in the spacelike region with respect to
the instrument and, as such, cannot be influenced by the
measurement process without violating the principles of the
special theory of relativity. Thus, both ways of explaining the
physical mechanism appear to be inconsistent, and this is exactly
what is called the paradox.

In arguing with the authors of~\cc{epr}, Bohr wrote in
article~\cc{bohr} that they incorrectly understood the notion of
"physical reality." According to Bohr, it is impossible for a
system with correlations to be considered as consisting of two
separate physical realities. Those correlations are physical
realities as well. Therefore, any measurement involving part of a
system should be considered as a measurement involving the whole
system. However, Bohr failed to back up his argument by providing
any clear physical interpretation. Of course, one can comfort
oneself with the fact that quantum mechanics is intrinsically not
interpretable in a commonsense way; nonetheless, dissatisfaction
with the situation still remains.

To circumvent this difficulty, Fock suggested considering that in
the quantum case the concept of a "state" should not be ascribed
objective meaning~\cc{fok}. Rather, it should be understood as
"information about the state." Here the question arises "is there
something objective we receive the information from?" In the
approach described in the second section of this article, it is
suggested to consider an elementary state to be that "something
objective."

In such a case, the physical phenomenon leading to the EPR paradox
can be given clear physical interpretation. Once an initial
particle has decayed, the physical system is characterized by
stable (zero) values of the observables $\s_{\nv}$ (a total spin
projection onto ${\nv}$). As a result, the values of observables
$\A_{\nv}$ and $\B_{\nv}$ (spin projections onto ${\nv}$ direction
for $A$ and $B$ particles, respectively) satisfy the following
relation:
 \beq{51}
  A_{\nv}+B_{\nv}=S_{\nv}=0.
  \eeq
In contrast to the quantum state, incompatible observables in the
elementary state may simultaneously have definite values, though
the latter cannot simultaneously be measured by means of a
classical instrument. In the experiment we can measure the
observable $\B_{\nv}$ for an arbitrary (although only one) ${\nv}$
direction, because, for different directions ${\nv}$ and ${\nv'}$
the observables $\B_{\nv}$ and $\B_{\nv'}$ are incompatible. By
virtue of equality~\rr{51}, we automatically measure the value of
the observable  $\A_{\nv}$ in such a measurement. This is
so-called indirect measurement. Thus, in such an approach, the EPR
paradox is trivially resolved.

One of the most frequently cited arguments in favor of physical
reality in the EPR sense being nonexistent is violation of Bell's
inequality~\cc{bell1,bell2}. Bell derived his inequality inspired
by the EPR proposition. After Bell numerous versions of analogous
inequalities were proposed. We shall focus on the version proposed
in~\cc{chsh}.

In this work the same physical system as in the present article
dealing with the EPR paradox is studied. A zero-spin particle is
considered that decays into two half-spin particles $A$ and $B$.
These particles fly apart to a large distance and are detected by
the instruments $D_a$ and $D_b$, respectively. The instrument
$D_a$ measures the spin projection of particle $A$ onto the $a$
direction, while  $D_b$ does the same over the $B$ particle.
Corresponding observables will be denoted by $\A$ and $\B$, while
the results of measurement are denoted by by $A_a$ and $B_b$.

Suppose that the state of the initial particle is characterized by
some physical reality that can be denoted by the parameter $\lll$.
This same parameter will be used to describe physical realities
that characterize decay products. Respectively, the results of
measurement of the observables $\A$ and $\B$ can be considered to
be the functions $A_a(\lll)$ and $B_b(\lll)$ of the
parameter~$\lll$.

Let the event distribution as a function of the parameter $\lll$
be characterized by the probabilistic measure $P(\lll)$:
 $$ \int P(d\lll)=1, \qquad 0\leq P(\lll)\leq 1. $$

 Introduce the correlation function  $E (a, b) $:
\beq{29}  E (a, b) = \int P(d\lll) \, A_a (\lll) \, B_b (\lll)
\eeq
 and consider the following combination:
  \bea {30}
M&=&|E (a, b) -E (a, b ') | + |E (a ', b) +E (a ', b ') | = \nn{}
 &=& \lt |\int P(d\lll)\,
A_a(\lll)\,[B_b(\lll)-B_{b'}(\lll)]\rt|+ \lt |\int P(d\lll)\,
A_{a'}(\lll)\, [B_b(\lll) +B_{b'}(\lll)]\rt|.
  \eea
For any directions $a$ and $b$
 \beq {31}
 A_a (\lll) = \pm1/2, \quad B_b (\lll) = \pm1/2.
 \eeq
 Therefore,
\bea{32} M & \le &\int
P(d\lll)\,[|A_a(\lll)|\,|B_b(\lll)-B_{b'}(\lll)|+ |A_{a '} (\lll)
| \, |B_b (\lll) +B_{b '} (\lll) |] =\nn{}
 &=&1/2 \int P(d\lll) \,
[|B_b(\lll)-B_{b'}(\lll)|+|B_b(\lll)+B_{b'}(\lll)|].\eea

Due to equalities~\rr{31}, one of the expressions
 \beq{33}
 |B_b(\lll) -B_{b'} (\lll) |, \qquad |B_b (\lll) + B_{b'}
(\lll) | \eeq
 for arbitrary $\lll$ is equal to zero, while the
other is equal to to unity. Note that both expressions contain the
same value of $\lll$.

Taking into account the property of expressions~\rr{33}, Bell's
inequality is deduced from~\rr{32},

 \beq{34} M \leq 1/2\int P(d\lll) =1/2. \eeq

 In standard quantum mechanics, the correlation
function is easily calculated to yield
 $$
 E (a, b) = -1/4\cos\theta_{ab},
 $$
 where $ \theta_{ab}$ is the angle between the directions $a$ and $b$.
 For the directions $a=0$, $b=\pi/8$, $a'=\pi/4$, and $b'=3\pi/8$
 we have
 $$ M=1/\sqrt{2}, $$
 which contradicts inequality~\rr{34}.

The results of the experiment~\cc{adr} are consistent with quantum
mechanical calculations and do not confirm Bell's inequality. As a
rule, these results are viewed as evidence that a quantum
mechanical system does not correspond to any physical reality that
would predetermine the results of the measurement.

Now we shall see that, in fact, application of probability theory
to quantum systems does not allow for such a derivation of Bell's
inequality

Since, in the quantum case, $\sss$-algebra and, respectively, the
probabilistic measure depend on the instrument used, the
replacement $P(d\lll)\to P_{\A\B}(d\vp)$ should be done in
definition~\rr{29}. And if we consider the correlation function
$E(a',b')$, then in~\rr{29} the replacement $P(d\lll)\to
P_{\A'\B'}(d\vp)$ is in order. Although in both cases the same
symbol $d\vp$ is used to denote an elementary volume in the space
of elementary states, it should be borne in mind that the set of
elementary states corresponding to $d\vp$ will be different. The
point is that these sets must be elements of $\sss$-àëãåáð. If the
observables $\A$ and $\B$ are incompatible with the observables
$\A'$ and $\B'$, òî $\sss$-algebra will be different. Moreover,
there is no physically justified $\sss$-algebra the subalgebras of
which could be identified with these algebras.

In addition, in the process of determining the value of the
correlation function $E(a,b)$ in the experiment we deal with a
random countable sample $\{\vp\}^{AB}_{\vx}$ selected out of the
total probability space $\Om(\vx)$, rather than with that complete
space itself. Finally, formula~\rr{29} must be replaced by

 $$ E(a, b) = \int_{\{\vp\}^{AB}_{\vx}}P_{\A\B}{d\vp)}\vp(\A\B). $$

Accordingly, formula~\rr{30} takes the following form:
 \begin{eqnarray*}
M&=& \lt |\int_{\{\vp\}^{AB}_{\vx}}P_{\A\B}{d\vp)}\, \vp(\A\B)-
\int_{\{\vp\}^{AB'}_{\vx}} P_{\A\B'}{d\vp)}\,\vp(\A\B')\rt|+ \nn{}
& +& \lt |\int_{\{\vp\}^{A'B}_{\vx}} P_{\A'\B}{d\vp)}\, \vp(\A'\B)
+ \int_{\{\vp\}^{A'B'}_{\vx}} P_{\A'\B'}{d\vp)}\, \vp(\A'\B')\rt|.
  \end{eqnarray*}

If the directions $a$ and $a'$ ($b$ and $b'$) are not parallel to
each other, then the observables $\A\B$, $\A\B'$, $\A'\B$, and
$\A'\B'$ are mutually incompatible. Hence, there is no universal
physically justified $\sss$-algebra, which would correspond to the
measurement of all these observables. Moreover, since the sets
$\{\vp\}^{AB}_{\vx}$, $\{\vp\}^{AB'}_{\vx}$,
$\{\vp\}^{A'B}_{\vx}$, and $\{\vp\}^{A'B'}_{\vx}$ are different
random samples from the continuum space $\Om(\vx)$, the
probability of their pairwise intersection is equal to zero.
Consequently, the probability of combinations of type~\rr{33}.
taking place is equal to zero as well. As a result, the arguments
that led to inequality~\rr{34} turn out to be incorrect for the
elementary states.

Thus, in a quantum case, the hypothesis of the existence of an
elementary state does not lead to Bell's inequality. Therefore,
numerous experimental verifications of this latter inequality
conducted so far become largely devoid of a theoretical basis.

\section{SINGLET STATE SIMULATION}

In the previous section, it was shown that the usually cited
proofs of Bell's inequality are substantially flawed, for they do
not account for a very important requirement of probability
theory: that the space of elementary events must be countable.
Otherwise, the very concept of probability itself loses a strict
mathematical sense. We shall see in this section that it is
possible to construct a macroscopic model in which Bell's
inequality will be violated.

To this end, let us construct a singlet state of a two-particle
system. A characteristic feature of a singlet state is that
equality~\rr{51} always holds during measurement of the spin
projection onto an arbitrary $\nv$ direction.

This equality assumes a hard correlation between the results of
measurements for the first and second particles regardless of how
far they are located from each other at the moment of measurement.
In the approach to follow in the present article, this assumes a
rigid correlation between the elementary states of the first and
second particles. In terms of the ESS, such a correlation is
easily noticed: the ESS for the second particle must be a negative
copy of the ESS for the first one. However, it is not clear in the
case of a singlet state what kind of the distribution in a set of
ESSs for the first particle should look like. It is simpler to pin
down such a distribution by using the MGS framework. Keeping in
mind the relation between the MGS and ESS mentioned earlier, it
can be expected that, in a singlet state, the MGS for the second
particle will be a mirror image of the MGS for the first one. By
this we mean that the layers in the MGS for the second particle
mimic those for the first one up to a change of orientation to the
opposite, i.e.,

\beq{17}
 \Rv_1^{(k)}+\Rv_2^{(k)}=0,
 \eeq
Here $\Rv_1^{(k)}$ and $\Rv_2^{(k)}$ are the orientation vectors
of the $k$-th layer for the first and second particles,
respectively. In addition, the functions $\ee^{(k)}(\rv)$ for
these particles satisfy the condition
$\ee^{(k)}_1(\rv)=\ee^{(k)}_2(-\rv)$. The latter condition, along
with equality \rr{17}, means that the numbers of active layers in
MGS for the first and second particles coincide.

From this it immediately follows that the equality
$\Rv^{(k)}_1\rv=-\Rv^{(k)}_2\rv$ holds for arbitrary $\rv$, which,
in turn, means that $j_1=-j_2$. That is, regardless of the
distance between particles, the spin projections onto the
arbitrary direction of $\rv$ registered at the moment of
measurement satisfy relation \rr{51}. This is the way the EPR
paradox is realized. The correlation in \rr{51} does not appear at
the moment of measurement; instead, it is a consequence of the MGS
ensemble structure that corresponds to a singlet state. This means
that the correlation appears at the moment the ensemble is formed.

Note that this state of affairs does not contradict Einstein's
remark on the incompleteness of quantum mechanics. In the model
proposed, the traditional mathematical apparatus of quantum
mechanics is supplemented by a novel concept --- that of an
elementary state.

Due to strong correlation between MGSs, it is sufficient to define
a set of elementary states for just one of two particles, say, for
the first one, so as to fix a set of elementary states
corresponding to a singlet quantum state.

We assume that the elementary states of the first particle are
described by MGSs that are elements of the set $\Upsilon$ and
satisfy the following conditions. The radius-vectors $\Rv^{(k)}$
in each MGS, which set a layer orientation, are randomly
distributed on the entire sphere $\RR$. The layer orientations in
the individual MGSs are distributed independently of each other.

Just like in a single-particle system, an elementary state
(elementary event) of a two-particle system cannot be ascribed any
probabilistic measure. Probability can only be attributed to
events that are elements of some $\sss$-algebra. This latter
algebra will be constructed in analogy to the single-particle
case.

Demand that the generatrices of $\sss$-algebra satisfy the
following conditions.
\begin{enumerate}
\item The generatrices form a subset of the set of two-particle
elementary states.
 \item Every two-particle elementary state is
described by two MGSs.
 \item In each of such MGSs, the conditions
 $\Rv^{(k')}_1\rv=-\Rv^{(k')}_2\rv$ and
$\ee^{(k')}_1(\rv)=\ee^{(k')}_2(-\rv)$ are satisfied for the
layers with the same number $k'$.
 \item $\rv_1$ and $\rv_2$ are
fixed; $\rv_1(\rv_2)$ is the direction in which the spin
projection of the first (second) particle is measured. These
vectors  $\rv_1$ and $\rv_2$ fix $\sss$-àëãåáðó. The remaining
conditions define elements of this algebra.
 \item The interval $d\ee$, $-1/2<\ee<+1/2$ is fixed for the
first particle. Either $\ee^{(k')}(\rv_1)\in d\ee$, ëèáî
$-\ee^{(k')}(\rv_1)\in d\ee$. Both options correspond to the same
subset (the same element of $\sss$-algebra).
 \item The number $k$ of active layer is fixed.
\item Each subset of elementary states for the first particle
consists of all MGSs that belong to $\Upsilon$ the layers of which
satisfy one of the following conditions:
  \bea{018}
 |\Rv_1^{(k')}\rv_1+\ee^{(k')}(\rv_1)|&>&1/2 \qquad k'=k,\\ \label{019}
|\Rv_1^{(k')}\rv_1+\ee^{(k')}(\rv_1)|&\leq&1/2 \qquad k'\leq k
\quad (k' \mbox{ fixed}).
 \eea
 These two conditions \rr{018} and \rr{019} correspond to
two different subsets.
 \item In the case of \rr{018}, the subsets differ in two more
parameters, $j_1$ and $j_2$: $j_1=+1$, if $\Rv_1^{(k)}\rv_1>0$;
 $j_1=-1$, if $\Rv_1^{(k)}\rv_1<0$; $j_2=+1$, if $\Rv_2^{(k)}\rv_2>0$;
 $j_2=-1$, if $\Rv_2^{(k)}\rv_2<0$.
 \end{enumerate}

Similarly to the single-particle case, it is easy to verify that
any elementary event from the set of singlet elementary states
belongs to any subset listed above irrespective of $\rv_1$ and
$\rv_2$.

Now, construct the probabilistic measure for each such subset
assuming that the random values of $\Rv_1$ are evenly distributed
on a sphere $\RR$, while the random values of $\ee$ are within the
interval $(-1/2,+1/2)$.

Let $\rv_1$, $\rv_2$, $d\ee$, $j_1$ and $j_2$ be fixed and $k'=1$.
Then, the probability for inequality \rr{018} to be satisfied is
described by the following expression:
\begin{eqnarray*}
& &P^{(1)}(\rv_1,\rv_2,\ee,j_1,j_2)d\ee\\
&=&d\ee N/2\int d\Rv \T[-j_2\Rv\rv_2]\Big[
\T[j_1(\Rv\rv_1+\ee)-1/2]+\T[j_1(\Rv\rv_1-\ee)-1/2]\Big]\\
&=&d\ee N/2\int d\Rv \T[-j_1j_2\Rv\rv_2]\Big[
\T[\Rv\rv_1+\ee-1/2]+\T[\Rv\rv_1-\ee-1/2]\Big].
\end{eqnarray*}
From the latter we obtain
 \beq{21}
P^{(1)}(\rv_1,\rv_2,\ee)=\sum_{j_1,j_2}P^{(1)}(\rv_1,\rv_2,\ee,j_1,j_2)
=N2\pi.
 \eeq
 The probability for inequality \rr{019} to be realized is
given as follows: \bea{20} \tilde P^{(1)}(\rv_1,\rv_2,\ee)d\ee
&=&d\ee N/2\int d\Rv \Big[
\T[1/2+\Rv\rv_1+\ee]\T[1/2-\Rv\rv_1-\ee]\nn &+&
\T[1/2+\Rv\rv_1-\ee]\T[1/2-\Rv\rv_1+\ee]\Big]=N2\pi d\ee. \eea

From\rr{21} and \rr{20}, we deduce
$$ N=(4\pi)^{-1}, \qquad \tilde P^{(1)}(\rv_1,\rv_2,\ee)=1/2.  $$

Furthermore, repeating the calculations carried out for a
single-particle system, we obtain that, for fixed $\rv_1$,
$\rv_2$, $d\ee$, $j_1$, and $j_2$ the probability for the active
layer to have the number $k$ is described by the formula
 \bea{22}
&\ P^{(k)}(\rv_1,\rv_2,\ee,j_1,j_2)\nn&=\frac 1{4\pi}\frac 1{2^k}
\int d\Rv \T[-j_1j_2\Rv\rv_2]\Big[
\T[\Rv\rv_1+\ee-1/2]+\T[\Rv\rv_1-\ee-1/2]\Big], \eea
 while the probability of having a number greater than $k$
 is given by
\beq{23} \tilde P^{(k)}(\rv_1,\rv_2,\ee)=2^{-k}. \eeq

Formulas \rr{22} and \rr{23} define the probabilistic measures for
the generatrices of $\sss$-algebra Proceeding with calculations
along the lines of \rr{13} and \rr{14}, we find the expression
 \beq{24}
 P(\rv_1,\rv_2,j_1,j_2)=\frac 14\Big(1-j_1j_2(\rv_1\rv_2)\Big), \eeq
which describes the probability of finding the spin projection of
the first particle onto the $\rv_1$ direction equal to $j_1/2$,
while for the second particle it is found to be $j_2/2$.

From formula \rr{24} we obtain for the correlation function
 \beq{25} E(\rv_1,\rv_2) =\frac 14\sum_{j_1,j_2}j_1j_2
P(\rv_1,\rv_2,j_1,j_2)=-\frac 14(\rv_1\rv_2). \eeq

 The distribution given in \rr{24} and the correlation
function in \rr{25}
  coincide with those obtained in the
standard quantum mechanics and violate Bell's inequality. For
$\rv_1=\rv_2$ and $j_1=j_2$, the right hand side of \rr{24}
vanishes. This corresponds to the EPR paradox.

 A few words on the locality issue in the model proposed are in order. The reason behind correlations between results
for the first and second particles appears at the time of shaping
out a singlet state rather than at the moment of measurement.
Specifically, this correlation is due to the fact that the numbers
of active layers in MGSs for both particles in each elementary
event coincide. Of course, it is assumed that each of the two
instruments can independently detect this active layer for its own
particle.

 In this article it is described in some detail how the
first instrument can find an active layer by using the functions
 $\ee^{(k')}(\rv)$, while for the second particle it is only
stated that its active layer has the same number. In order to
render the measurement processes for the first and second
particles completely independent, we can take the following
approach. Starting from the MGS (and the functions
$\ee^{(k')}(\rv)$) studied for the first particle, we construct a
corresponding ESS according to the technique given above. For the
second particle, the ESS is constructed as a negative copy of the
ESS for the first one. All these operations are related to the
process of shaping of the state studied, rather than that of a
measurement. By using the ESSs of individual particles shaped in
this manner, each of the instruments can independently measure the
spin projection onto a desired direction.

 Such a scheme of measurement is not convenient for
computer simulation. More suitable is a scheme that simulates the
experiment with a so-called delayed choice. It can be outlined as
follows. In the elementary event considered, the number of active
layer is fixed for the first particle along, with measurement of
the spin projection onto the $\rv_1$ direction. For the second
particle in the same elementary event, the inequalities
$\Rv_2\rv_2>0$ and $\Rv_2\rv_2<0$ are verified for a sufficiently
large number of layers and the corresponding values of
 $j_2$ with the
indicated number of layer are fixed. The accuracy of the final
result depends on the number of layers checked. Since the
contribution to this result rapidly decreases with increasing
number of active layers, this accuracy promptly increases with an
increase in the number of layers checked.

 Once the observational part of the experiment is
over, the measurement data processing stage is set to commence. If
the goal of the experiment is to reveal certain correlations
between results of measurements for the first and second
particles, then the latter must be stored in one place. In this
sense this stage will be necessarily nonlocal. However, this is
not a distinctive feature of the quantum experiment; the same is
true of the classical one.

 In the case at hand, only those values of $j_2$ obtained
for different layers should be chosen while processing the data,
which correspond to the layer whose number coincides with that of
an active layer found by the observation with the first particle.
This procedure can be viewed as a simulation of the coincidence
process that is necessary in such a type of experiment. Thus, the
entire nonlocality is exclusively associated with processing of
the measurement data.

 In the process of developing the computer simulation procedure, it became clear that a finer partition of the set
of elementary events into subsets of probabilistic events is
possible. Namely, events corresponding to $\ee(\rv)\in d\ee$ and
 $-\ee(\rv)\in d\ee$ can be regarded as belonging
to different elements of $\sss$-algebra. Such a $\sss$-algebra is
convenient for computer simulation, while for analytical
evaluations it is less favorable.

\section{CONCLUSIONS}

The procedure described can be considered as a working model of
quantum measurement, rather than a mere imitation of it.
Consequently, this model is quite suitable for conducting
experiments with the aim of verifying any statements about quantum
systems

Like any particular model, it is not very well suited for
substantiating {\it positive statements}, i.e., for justifying the
fact that the statement expressed is unconditionally correct. On
the other hand, it is perfectly suited for {\it negative
statements}, i.e., for disproving that a certain statement is
correct.

In particular, this model refutes the frequently adduced statement
that experimentally observed violation of Bell's inequality proves
the absence of a local physical reality that is a source for
quantum phenomena.

The model proposed might appear to be very useful for conducting
experiments with so-called quantum teleportation (see, for
example, \cc{bpw}). Hopefully, it will help demystify this
physical phenomenon. For more detail, refer to \cc{slav3}.

On the other hand, for the purposes of quantum cryptography, the
model proposed is only partially suitable. It can be used for
experimental verification of any statements made in this field.
However, this model is not suitable for practical use in quantum
cryptography. The issue here is that the latter is based on the
existence in quantum systems of observables that can in no way be
simultaneously measured, whereas in the model proposed there are
observables that are incompatible only with respect to a
particular measurement process.

\end{document}